Torgom Yezekyan [1], Roza Gabrielyan [2], Sergey I. Bozhevolnyi [3]

# Quasi-bound states in the continuum in finite waveguide grating couplers

**Abstract:** Finite-size effects occurring for quasi-bound states in the continuum (qBIC) formed in symmetric and asymmetric all-dielectric waveguide grating couplers are investigated using numerical simulations for different configuration parameters. We find that the beam size plays a crucial role in the formation of additional qBIC resonances originating in forbidden for plane-wave incidence BIC resonances, whose manifestation is strongly influenced by the beam divergence. Another parameter is the strength of Bragg reflection, determining the spatial extension of a distributed Bragg resonator (DBR): in configurations with strong Bragg reflection, the excited and Bragg reflected counterpropagating waveguide modes are confined within the footprint of the incident beam. Conversely, the DBR formed in configurations with weak Bragg reflection extends well beyond the incident beam footprint, resulting in high sensitivity of the system response to the grating width and incident beam position. We believe that our findings provide valuable insights for optimizing photonic devices that leverage qBIC resonances in finite-sized structures under realistic excitation conditions.

[1] POLIMA—*Center for Polariton-driven Light–Matter Interactions, University of Southern Denmark, Campusvej 55, 5230, Odense M, Denmark*
[2] *Institute of Physics, Yerevan State University, Alex Manoogian 1, 0025, Yerevan, Armenia*
[3] *Center for Nanooptics, University of Southern Denmark, Campusvej 55, 5230, Odense M, Denmark*

## Introduction

In 1929, Von Neumann and Wigner found localized solutions of the Schrodinger equation for a single particle, where energies of these eigenstates lie in the continuum of the eigenvalue state solutions, the so-called Bound States in the Continuum (BICs) [1, 2]. Over time, the particular interest in investigating these embedded states led to their experimental realization [3-6]. BICs are characterized as non-radiating resonant modes in open non-Hermitian systems, where their energy remains perfectly localized with an infinite lifetime, despite residing within the continuous spectrum of radiative waves. These modes being decoupled from radiating channels, prevent propagation outside the system. According to theory, BICs can give rise to infinite quality factors (Q-factor), with several types of excitation mechanisms based on their physical origins [7-9]. The Friedrich-Wintgen BIC mechanism, resulting from destructive interference between two resonant leaky modes in the same cavity, is one of the most studied methods [10-13]. For periodic photonic structures this type can be realized, if the mode frequency is below the frequency of excitation under normal light incidence and the structure is invariant under the time-reversal, in-plane inversion.

On the other hand, any perturbation in the ideal system, results in the collapse of BIC to a Fano resonance with a finite lifetime, a regime that is known as quasi-BIC (qBIC) and characterized by its quadratic dependence of Q-factor on the perturbation [9]. Another interesting regime is the so-called "near BIC," at which very high $Q$ factors are still attainable in the vicinity of the BIC [14]. A central advantage of the near-BIC regime is that, unlike regular guided modes below the light line, the modes in this regime can be excited by free propagating plane waves.

Investigations of all these resonances have rapidly accumulated to constitute a cornerstone in modern photonics, offering unique opportunities for designing high-Q resonant systems with applications spanning sensing, lasing, and nonlinear optics [6, 8, 15-19]. While much progress has been made in understanding BICs and qBICs in idealized, infinite structures, real-world implementations that inevitably deviate from these theoretical paradigms are practically unexplored. Two critical factors that influence the behavior of these resonances are the finite size of photonic structures and the finite spatial extension of excitation beams. Investigating influence of these factors on resonant interactions in photonic structures is essential to bridge the gap between theoretical predictions and experimental realizations.

In this work, we investigate all-dielectric symmetric and asymmetric waveguide grating couplers based on the waveguide configuration investigated previously [13] but consider here the grating couplers that are finite in size and excited with (normally incident) finite-size Gaussian beams. The essential physics remains however the same: the resonant excitation of waveguide modes under normal light incidence, taking place at the first diffraction order, is intrinsically linked to the occurrence of Bragg reflection of counterpropagating waveguide modes, taking place at the second diffraction order, and results thereby in the formation of a distributed Bragg resonator (DBR) along with BIC and qBIC resonances [13]. We study the finite-size effects by first varying the excitation Gaussian beam radius to determine the smallest beam radius that ensures a practically complete transmission extinction of a normally incident beam with reasonably large gratings. Then, by varying the grating width and position of the excitation beam for the two gratings with different parameters that influence the Bragg reflection strength and thereby the spatial extension of the DBR, we explicitly reveal the importance of the grating width and Gaussian beam radius for manifestation of qBIC resonances: weaker Bragg reflection requires wider gratings and incident excitation beams for maintaining close to complete transmission extinction of the incident beam. In the opposite case of strong Bragg reflection, the excited and Bragg reflected counterpropagating waveguide modes are confined within the footprint of the incident beam. Our findings provide crucial insights that bridge the gap between theoretical models of infinite structures and practicalities of experimental work.

## Results and discussion

The formal design of the photonic structure we study is shown on Figure 1 (a), which consists of a planar waveguide with a grating (periodic dielectric stripes) of finite lateral size on top. In our recent work we thoroughly investigated occurrence of BIC and qBIC resonances in similar structures with infinite gratings, under a plane wave incidence [13]. Despite the differences, the physical mechanisms involved in the resonance formation are the same, i.e., the resonant excitation of waveguide modes with the grating in the first diffraction order and the inherent Bragg reflection of those counterpropagating waveguide modes in the second diffraction order.

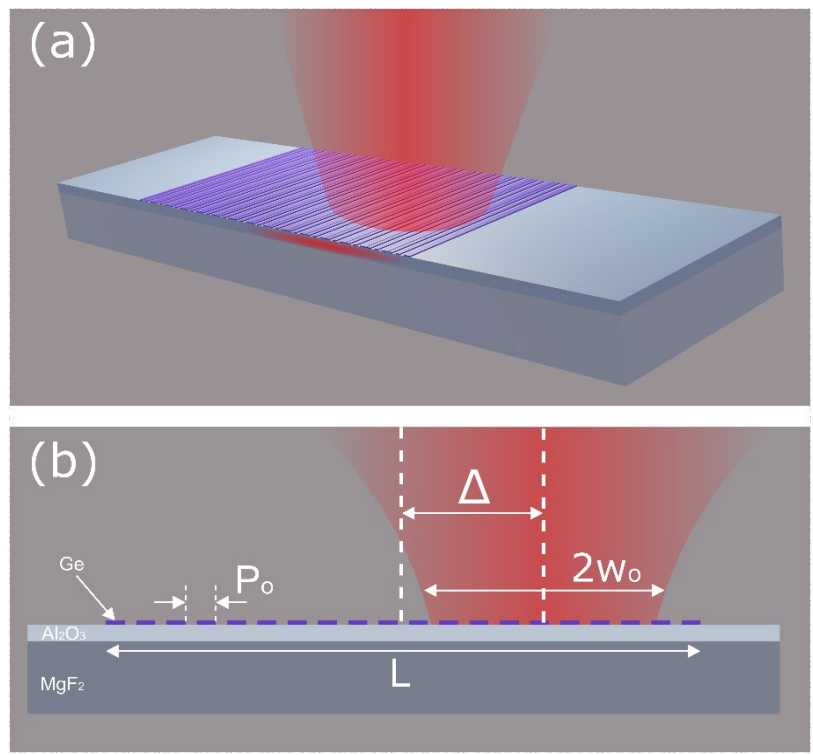

Figure 1. Schematic of the thin-film waveguide configuration with a grating coupler of finite size on its top (a). Off-center excitation of waveguide modes with a Gaussian beam characterized with the waist radius $w_0$ and beam offset $\Delta$ by a grating coupler with the period $P_0$ and width $L$ (b).

In our previous work we have shown that, from the viewpoint of interplay between the diffractive coupling to counterpropagating waveguide modes and their Bragg reflection within the same grating, there exists a conceptual difference when considering symmetric and asymmetric gratings [13]. Therefore, we consider here two grating couplers: a symmetric grating with periodically distributed identical dielectric stripes and an asymmetric one, in which every sec-ond strip is perturbed in width. In the case of symmetric gratings, the waveguide mode wavelength ($\lambda_w$) should be equal to the grating period $P_o$ for enabling the mode excitation in the first diffraction order, while the DBR condition implies $2P_o = m\lambda_w$, where $m$ is the order of Bragg reflection. These two conditions are satisfied only in the case of second order Bragg reflection. Asymmetric grating, on the other hand, consists of periodic dielectric stripes with period of $P_o/2$, while the supercell period is kept the same ($P_o$) to maintain the first-order excitation of waveguide modes at the same wavelengths. Thus, in this case, we still rely on the excitation in the first diffraction order ($\lambda_w = P_o$), whereas the DBR condition is modified due to the existence of a twice smaller period of (unperturbed) individual stripes in the system, entailing the following condition: $P_o = m\lambda_w$. Here one ends up with a distinctly different physics, since the mode excitation occurs only due to the stripe width perturbation, while the Bragg reflection is realized in the first order of a strong grating of unperturbed stripes.

To understand the influence of the beam size on the waveguide mode excitation and consequent formation of BIC and qBIC resonances we begin our analysis with a symmetric grating coupler of the period $P_o = 1200$ nm and lateral size of $L \approx 100\ \mu m$ (Figure 1 (b); the structure is infinite along the grating stripes). A 1.2 $\mu m$-thin film of $Al_2O_3$ on a 500 $\mu m$-thick $MgF_2$ substrate serves as a waveguiding configuration, while the grating is formed by 50-nm-high and 880-nm-wide germanium stripes (Figure 2 (a)). All the parameters are chosen to be consistent with the ones from our previous investigation [13]. The simulations are performed for 2D models in the commercially available finite element software COMSOL Multiphysics 6.2. Interpolated values are used for the permittivity of $Al_2O_3$ [20], $MgF_2$ [21], Ge [22]. The transmission spectra of a normally incident TE-polarized Gaussian beam calculated for different beam waist radii ($w_o$) are shown in Figure 2(a) for the beam being centered (offset $\Delta = 0$). For the reference, the transmission spectrum of a normally incident plane wave is also presented alongside (empty circles in Figure 2(a)). It is seen that, for relatively wide Gaussian beams ($w_o = 40$ and $80\ \mu m$), there is practically no difference in the transmission spectra when compared with the plane wave incidence: the transmission is completely extinct at the resonant wavelengths corresponding to the excitation of fundamental and first transverse electric (TE) modes [13], and these wavelengths remain practically the same. The transmission spectrum becomes completely different for the small beam radius ($w_o = 5$ μm, the blue curve in Figure 2(a)). In this case, one observes the occurrence of an additional resonance minimum at the long-wavelength side of the (shifted to a shorter wavelength) minimum corresponding to that observed for the plane-wave and wide-beam normal incidences. Note that the new long-wavelength minimum for the fundamental mode is outside of the considered wavelength interval.

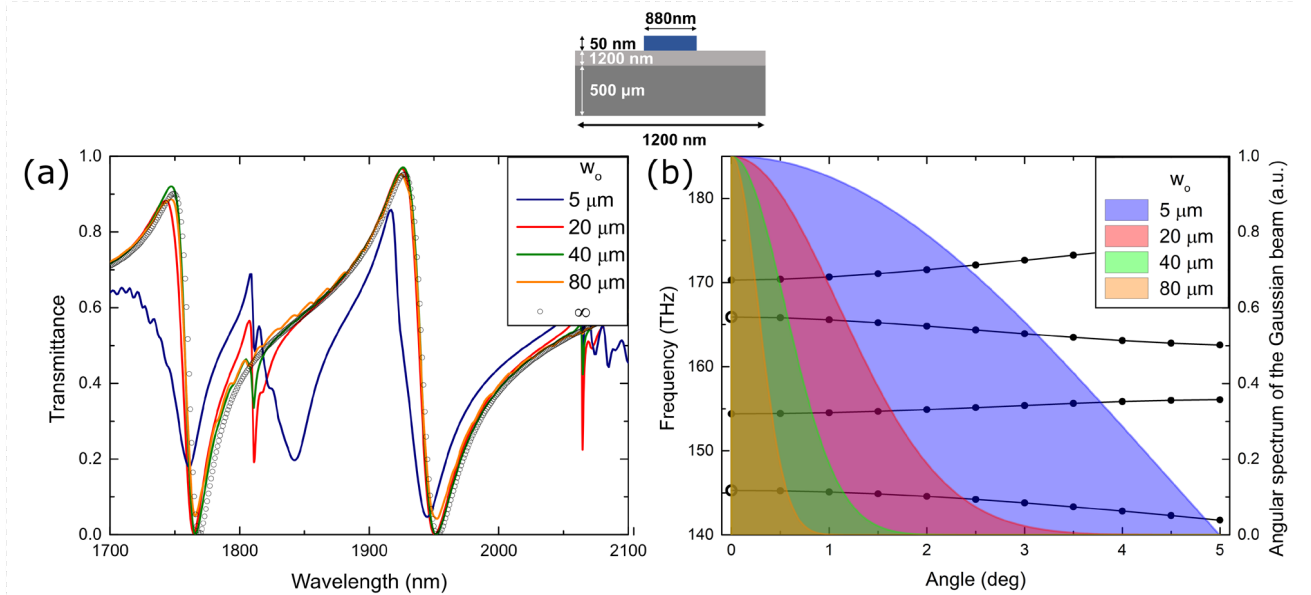

Figure 2. Transmission spectra for normal Gaussian beam incidence at the 100-$\mu m$-wide symmetric grating coupler calculated for different beam waist radii $w_o$ (a). Dispersion curves calculated for the normal plane-wave incidence at the infinitely wide grating coupler [13] along with angular (spatial frequency) spectra of Gaussian beams with different beam waist radii corresponding to the selected transmission calculations (b).

The occurrence of additional transmission minima for the narrow-beam normal incidence is apparently similar (although with noticeable differences in the perceived Q-factors) to what was observed for off-normal plane-wave incidence in our previous work: that has also resulted in the occurrence of additional minima at the long-wavelength side and blueshifts of the minima related to the normal plane-wave incidence [13]. The physical mechanisms at play are also similar. Indeed, the spatial frequency spectra of finite-size Gaussian beams contain non-zero spatial frequencies corresponding to non-zero angles of incidence of plane-wave components originating from the plane-wave decomposition of Gaussian beams, with narrower beams resulting in wider spatial frequency spectra and larger incident angles (Figure 2(b)). For the narrow (5-$\mu m$-wide) incident beam, the (non-zero) plane-wave components associated with the off-normal incidence destroy partially the BIC symmetry protection and thereby result in the occurrence of the additional transmission minima at the (longer) wavelengths (Figure 2(a)) corresponding to the lower, BIC related, dispersion curves calculated for the normal plane-wave incidence at the infinite grating coupler (Figure 2(b)). The additional minimum observed in Figure 2(a) is understandably broader than those calculated for off-normal incident plane waves [13] because it is caused by the angular spectrum of incident plane-wave components as opposed to one specific off-normal incident wave. The pronounced (for the narrow beam) blueshifts of the original minima associated with the qBIC resonances [13] correspond to the qBIC-related dispersion curves tending to higher frequencies (shorter wavelengths) away from the normal incidence (Figure 2). Note that, for different incident beams, there might appear different minimum-like spectral features (at longer wavelengths) related to different plane-wave components within the angular spread of the incident beam spatial frequency spectrum. In fact, any finite-sized normally incident beam generates off-normal incident plane-wave components, that should result in more (for narrower beams) or less (for wider beams) pronounced additional spectral features at the long-wavelength side of the original, qBIC related, resonance that becomes progressively (for narrower beams) wider and more shifted to shorter wavelengths, features that can also be inferred from the transmission spectra and dispersion curves shown in Figure 2.

At this point, we would like to comment on the terminology used to describe the considered physical phenomena. For normally incident (on infinite gratings couplers) plane waves, the resonant excitation of counterpropagating waveguide modes is intrinsically linked to the DBR formation and occurrence of the band gap revealed with two dispersion branches starting off the associated BIC and qBIC resonances (Figure 2(b)). The off-axis points of the BIC-related dispersion is accessible with off-normal plane-wave incidence, resulting in narrow resonance-like transmission minima [13]. These resonances are typically referred to as guided-mode (GM) resonances [23] or near-BIC resonances [14]. However, at off-normal incidence, the Bragg reflection of the excited waveguide mode can no longer be guaranteed since the Bragg condition is no longer satisfied. In the absence of the Bragg reflection, no DBR is formed, and the transmission minimum should simply be related to the phase-matched excitation of a single (propagating away from the excitation domain) waveguide mode. This transmission minimum is therefore a Fano interference phenomenon that should not be confused with any of the resonances [13], because the occurrence of resonance implies the existence of resonant field configuration, which in our case can only be generated by the DBR. At the same time, the Bragg reflection has a certain bandwidth manifesting itself, in our case, via the occurrence of the band gap in dispersion diagrams. If the excited waveguide mode is within the spatial frequency bandwidth of the Bragg reflection, the DBR can be formed, and the corresponding transmission minimum can be associated with the GM or near-BIC resonances. The same reasoning should also be applied to the qBIC-related dispersion branch: transmission minima observed at off-normal incidence can be associated with qBIC resonances if the excited waveguide modes are within the spatial frequency bandwidth of the Bragg reflection.

Returning to the main subject of finite-size effects, another important parameter is associated with the strength of Bragg reflection determining the DBR spatial extension, i.e., the effective DBR width, that indicates how tight the excited and Bragg-reflected (counterpropagating) waveguide modes are bound to the lateral excitation region, i.e., to the width of normally incident Gaussian beam. We consider the 40-$\mu m$-diameter Gaussian beam ($w_0 = 20\ \mu m$) centered with respect to the 100-$\mu m$-wide grating coupler as a reference configuration, whose transmission spectrum fairly replicates that for the plane wave incidence on an infinite grating (Figure 2 (a)). The intention is to evaluate the effective DBR width by following the evolution of the transmission minimum related to the excitation of waveguide mode when varying the grating width and position of the excitation beam away from the reference configuration. We expect to find that the grating couplers with different parameters would influence the Bragg reflection strength and thus the spatial extension of the DBR, whose changes would be reflected in modifications of dependences of the transmission minimum on grating parameters. It is anticipated that, for strong Bragg reflection, the excited and Bragg-reflected counterpropagating modes are confined within the footprint of the incident beam, while, for weak Bragg reflection, the DBR extends well beyond the incident beam. It is also expected that larger grating ridge widths lead to larger scattering strengths for any diffraction order.

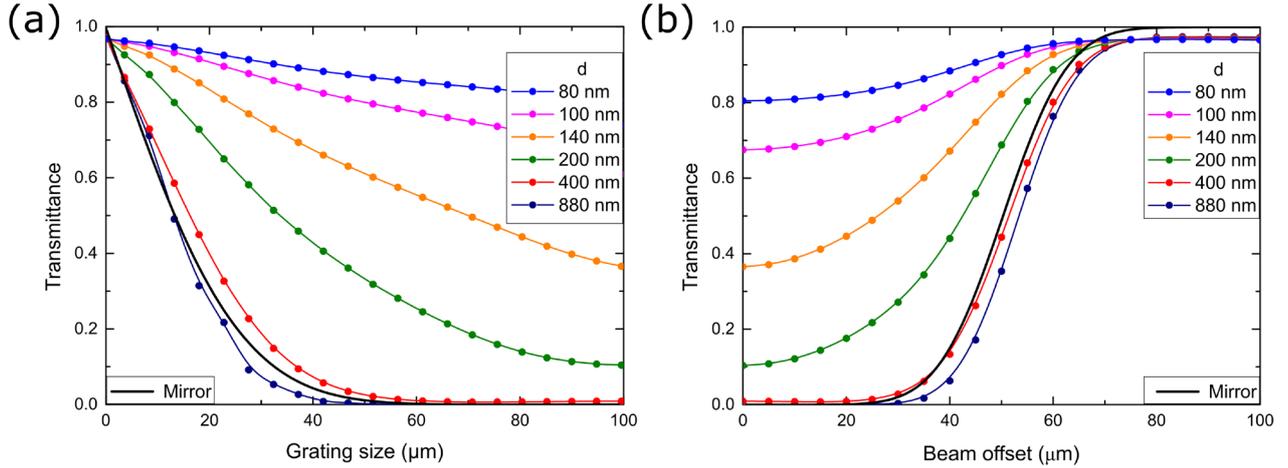

Figure 3. Transmission of symmetric grating couplers for normal 20-*µm*-wide Gaussian beam incidence calculated for different ridge widths *d* as a function of the grating size *L* (a) and beam offset *Δ* (b). Black solid line in both panels corresponds to the transmittance of the perfect reflector equal in width to the considered grating coupler illuminated with the considered Gaussian beam.

The dependences of the transmission minimum (resonant transmittance) on the grating size *L* show that, for relatively large grating ridge widths of 400 and 880 nm, the grating size can be decreased down to 50 *µm* without deterioration of very low transmittance at resonance (Figure 3(a)). Even the 40-*µm*-wide grating, i.e., of the same size as the incident Gaussian beam, exhibits low (< 10%) transmittance, unequivocally indicating that the DBR is within a footprint of the incident beam. To further support this assertion, we have modelled the grating coupler as a perfect reflector, whose transmission is only related to those parts of the 40-*µm*-diameter incident Gaussian beam that are not incident on the grating. The transmittance, in this case, can simply be evaluated by integrating the beam intensity profile over beam parts extending beyond the grating. The resulting transmittance dependence on the width of a perfect reflector (black solid line in Figure 3(a)) is indeed very close to those calculated for relatively large grating ridge widths of 400 and 880 nm. For smaller ridge widths, even the largest considered grating width of 100 *µm* is not sufficient to reach the transmittance level that no longer depends on further grating width increase (Figure 2 (a)), thus indicating that the Bragg reflection becomes too weak and the DBR extension too large to efficiently suppress the incident 40-*µm*-wide beam transmission within the 100-*µm*-wide grating.

Here it should be borne in mind that, in the ideal case of normal plane-wave incidence on lossless infinitely extended waveguide gratings, the qBIC resonance transmission would be zero for any grating parameter, because the outcoupled (by the grating) radiation would interfere distractively (in transmission) or constructively (in reflection) with the incident one. For a finite-size beam normally incident on a finite-size grating, the transmission is never zero, because the Bragg reflection associated with the exponential decay of the incident field allows some part of the incident waveguide mode propagate beyond both the incident beam and grating coupler areas, escaping thereby the Bragg reflection and destructive interference for outcoupled waves. For weak gratings, an appreciable part of the excited waveguide modes reaches the grating terminations (escaping thereby the Bragg reflection) and continues their propagation in unperturbed waveguide without leakage into either of the neighbor media. Note that, even for infinite grating couplers, the resonant transmission of finite-size beams increases for weaker gratings, because the DBR width becomes progressively larger than the incident beam width. Consequently, the waveguide modes outcoupled by the grating outside the incident beam area can no longer interfere with the incident beam, cancelling each other in transmission, and the transmission increases.

One would expect that similar conclusions could be surmised from the transmittance dependencies on the incident beam displacement, offset *Δ*, from the grating center (Figure 3(b)). Indeed, it is seen that, for relatively large grating ridge widths of 400 and 880 nm resulting in the strong Bragg reflection and DBR confinement (within the incident beam footprint), the transmittance offset dependencies follow closely that of a perfect 100-*µm*-wide reflector, increasing from very small values for small offsets to near unity for beams incident away from the grating coupler. Note that the incident beam is centered at the grating edge for the offset of 50 *µm*, touching the edge from within and outside the grating at 30 and 70 *µm*, respectively. For weaker gratings and thus larger DBR extensions, the influence of the grating edge causing the increase in transmission is observed at progressively smaller offsets (Figure 3(b)).

The physical mechanisms involved are the same as those considered above when discussing the influence of the grating size on the transmission (Figure 3(a)). It would also be reasonable to comment that these dependencies are, at least to some extent, complementary. The most important conclusion of the discussed simulations and their considerations is that sufficiently strong waveguide grating couplers do not have to be larger than the incident beam size to achieve efficient suppression (in transmission) of the incident beam.

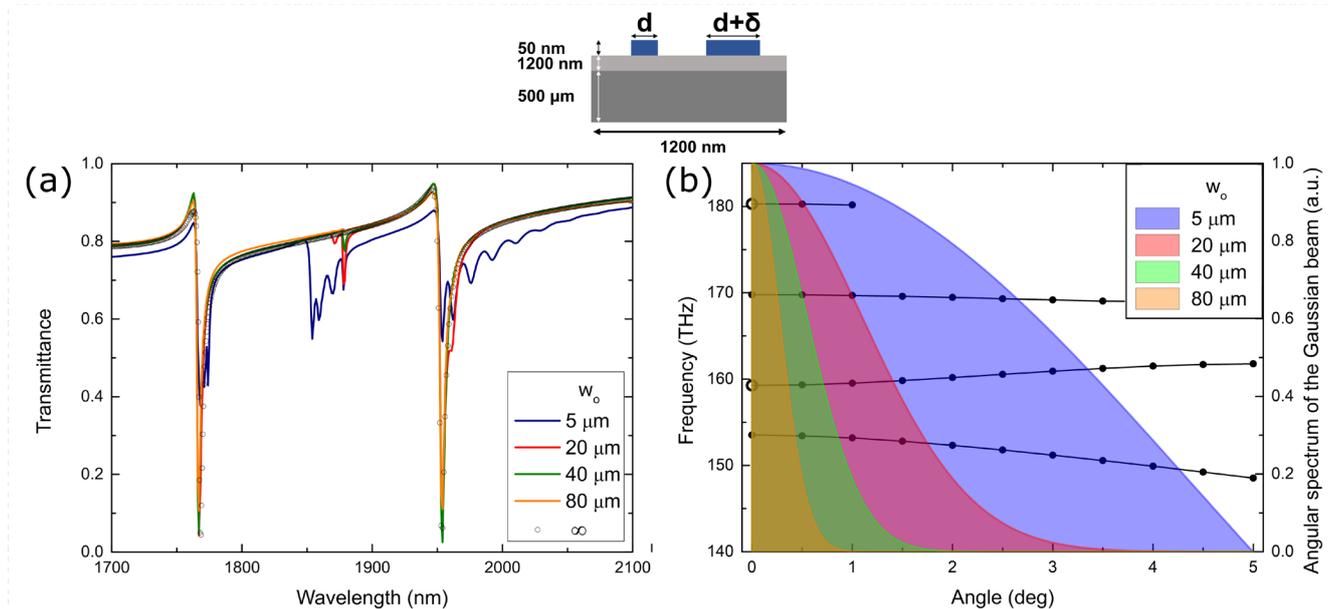

Figure 4. Transmission spectra for normal Gaussian beam incidence at the 100-$\mu$m-wide asymmetric grating coupler for different beam waist radii $w_0$ (a). Dispersion curves calculated for the normal plane-wave incidence at the infinite grating coupler [13] along with angular (spatial frequency) spectra of Gaussian beams with different beam waist radii corresponding to the selected transmission calculations (b). The nominal ridge width $d$ is 150 nm, while the perturbation $\delta$ is 80 nm.

We can now turn our attention to finite-size effects observed with the asymmetric grating, whose parameters are again consistent with those from our previous investigation [13], implying the same supercell period $P_0$ = 1200 nm and grating width L ≈ 100 $\mu$m. As discussed above, in this case, the mode excitation occurs only due to the stripe width perturbation, while the Bragg reflection is realized in the first order of a strong grating of unperturbed stripes. The transmission minima for a normal plane-wave incidence on an infinitely wide grating are significantly narrower in this case than those observed for the symmetric grating (cf. empty circles in Figures 2(a) and 4(a)), a remarkable difference that was attributed to relatively weak radiation loss (i.e., in- and outcoupling of excited waveguide modes) for the asymmetric grating [13]. The influence of finite sizes of incident Gaussian beams on the transmission spectra is somewhat similar to the symmetric grating case: narrow Gaussian beams open access to near BIC resonances, although the occurrence of new transmission minima is mirrored (Figure 4(a)). Contrary to the symmetric grating case, additional (associated with BIC dispersion branches) minima for narrow beams occur at the short-wavelength side of those observed for the plane-wave and wide-beam normal incidences, because the corresponding BICs are formed on the high-energy (short-wavelength) side of the band gap (Figure 4(b)). Note that the new short-wavelength minimum for the first-order mode is outside of the considered wavelength interval.

The dependences of the transmission minimum (resonant transmittance) on the grating width $L$ calculated for asymmetric gratings with different perturbations reveal rather strong influence of the perturbation parameter $\delta$ (Figure 5(a)), indicating that the spatial DBR extension depends not only on the Bragg reflection strength that, being determined by the unperturbed grating, should not vary much for small perturbations ($\delta < d$). Apparently, the radiative coupling (in-coupling and outcoupling) of waveguide modes, being strongly dependent on the perturbation parameter (radiative coupling vanishes for small perturbations), plays also an important role. Both effects add up constructively, because larger perturbations result in larger average size of ridges (responsible for the first-order Bragg reflection) and more pronounced supercell gratings enabling the first-order radiation coupling. Thus, the asymmetric grating made of 150 and 320-nm-wide ridges, i.e., having a very large perturbation $\delta$ = 320 nm, results in the DBR confined within the incident beam width, which in turn leads to the transmittance dependence being very close to that of the perfect reflector and thereby to that of the symmetric grating with the largest considered ridge width $d$ = 880 nm (cf.

Figures 3(a) and 5(a)). Note that the dependence of perfect reflector transmittance on the grating (reflector) width is naturally identical in both cases.

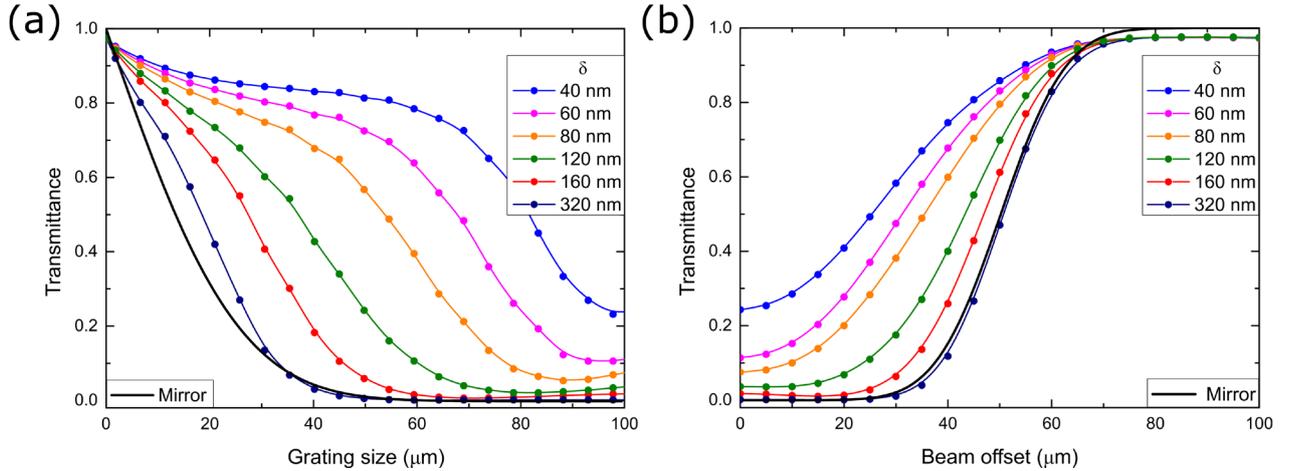

Figure 5. Transmission of asymmetric grating couplers for normal 20-$\mu$m-wide Gaussian beam incidence calculated for different ridge perturbations $\delta$ as a function of the grating size $L$ (a) and beam offset $\Delta$ (b). Black solid line in both panels corresponds to the transmittance of the perfect reflector equal in width to the considered grating coupler illuminated with the considered Gaussian beam.

The above conclusions are additionally supported by the behavior of the transmittance dependencies on the incident beam displacement, offset $\Delta$, from the grating center (Figure 5(b)): for the largest considered ridge perturbation $\delta$ = 320 nm, the transmittance dependence is very close to that of the perfect reflector, unequivocally indicating that the DBR is confined within the incident beam footprint. Conversely, for asymmetric gratings with relatively small perturbations, the DBR extends progressively farther and beyond the beam footprint, causing the increase in transmission being observed at progressively smaller offsets (Figure 5(b)). The physical mechanisms responsible for the influence of both the grating size and beam offset on the resonant transmission are the same as those considered above for the case of symmetric gratings. Also, for asymmetric gratings one finds that the transmittance dependencies on the grating size and beam offset are complementary. This complementarity can in both cases be accounted for, at least partially, by the choice of 100 $\mu$m for both the largest grating size and largest offset. For the 40-$\mu$m-diameter incident beam, this choice implies completely reversed situations at both ends of the interval, within which the grating size and offset are varied.

The considered waveguide grating coupler configurations are different in underlying physics and transmission spectra. At the same time, while the widths of transmission minima found for plane-wave incidence on infinitely extended gratings differ by one order of magnitude [13], the dispersion curves indicate similar band gaps (cf. Figures 2(b) and 4(b)). Considering the finite-size effects studied in this work, we arrived at similar important practical conclusions in a sense that, for both configurations, 40-$\mu$m-diameter incident Gaussian beams are found wide enough to ensure nearly complete extinction of the transmission with 40-$\mu$m-wide waveguide grating couplers. This similarity indicates that the most important physical parameter is the strength of resonant Bragg reflection that determines both the size of the band gap in the dispersion curves and the manifestation of finite-size effects, although the strength of the radiative coupling plays a certain role. Fundamentally, we demonstrated that the waveguide grating couplers can be engineered to exhibit tightly confined interaction between incident radiation and excited waveguide modes, i.e., the DBR limited to the incident beam footprint.

## Conclusions

Summarizing, the focus of the present work was to study finite-size effects in all-dielectric symmetric and asymmetric waveguide grating couplers for normal light incidence by considering the waveguide configuration investigated previously [13]. We have investigated the finite-size effects by first varying the excitation Gaussian beam radius to determine the smallest beam radius that ensures a practically complete transmission extinction of a normally incident beam with reasonably large gratings. Then, by varying the grating width and position of the excitation beam for the two considered gratings we have explicitly revealed the extent of interaction between incident radiation and excited waveguide modes, i.e., the spatial DBR extension, and studied the influence of grating parameters on the interaction spatial confinement, i.e., the DBR extension. We have identified the strength of Bragg reflection as the most important parameter determining the spatial DBR extension: in configurations with strong Bragg reflection, the excited and Bragg reflected counterpropagating waveguide modes are confined within the footprint of the incident beam. Conversely, the DBR formed in configurations with weak Bragg reflection extends well beyond the incident beam footprint, resulting in high sensitivity of the system response to the grating width and incident beam position. Importantly, we have demonstrated that the waveguide grating couplers can be engineered to feature the DBR limited to the incident beam footprint, an important design objective enabling the footprint minimization of the corresponding photonic components (waveguide couplers or narrowband wavelength filters). Overall, we believe that our findings shed light on important physical mechanisms at play for qBIC resonances in finite-size waveguide gratings, providing valuable insights for optimizing waveguide-grating-based photonic devices that leverage qBIC resonances in finite-sized structures under realistic excitation conditions.

**Acknowledgment**:

**Research funding**: TY acknowledges the support from the Center for Polariton-driven Light-Matter Interactions (PO-LIMA) funded by the Danish National Research Foundation (Project No. DNRF165). RG acknowledges the support from Scientific Research Grant through the State Committee of Science of Ministry of Education, Science, Culture and Sports of Armenia (Project No. 23RL-2A034).

**Conflict of interest statement**: The authors declare no conflicts of interest regarding this article.

# Bibliography

1.  C.W. Hsu, B. Zhen, J. Lee, S.-L. Chua, S.G. Johnson, J.D. Joannopoulos, and M. Soljačić, *Observation of trapped light within the radiation continuum,* Nature **499**(7457), 188 (2013).
2.  J. Von Neumann and E. Wigner, *On some peculiar discrete eigenvalues,* Phys. Z **30**, 465 (1929).
3.  C.W. Hsu, B. Zhen, A. Stone, J. Joannopoulos, and M. Soljačić, *Nature Rev,* Mater **1**, 16048 (2016).
4.  K. Koshelev, A. Bogdanov, and Y. Kivshar, *Meta-optics and bound states in the continuum,* Science Bulletin **64**(12), 836 (2019).
5.  D. Marinica, A. Borisov, and S. Shabanov, *Bound states in the continuum in photonics,* Physical review letters **100**(18), 183902 (2008).
6.  Z. Sadrieva, M. Belyakov, M. Balezin, P. Kapitanova, E. Nenasheva, A. Sadreev, and A. Bogdanov, *Experimental observation of a symmetry-protected bound state in the continuum in a chain of dielectric disks,* Physical Review A **99**(5), 053804 (2019).
7.  E.A. Bezus, D.A. Bykov, and L.L. Doskolovich, *Bound states in the continuum and high-Q resonances supported by a dielectric ridge on a slab waveguide,* Photonics Research **6**(11), 1084 (2018).
8.  M. Kang, T. Liu, C. Chan, and M. Xiao, *Applications of bound states in the continuum in photonics,* Nature Reviews Physics **5**(11), 659 (2023).
9.  K. Koshelev, S. Lepeshov, M. Liu, A. Bogdanov, and Y. Kivshar, *Asymmetric metasurfaces with high-Q resonances governed by bound states in the continuum,* Physical review letters **121**(19), 193903 (2018).
10. M. Amrani, S. Khattou, E.H. El Boudouti, A. Talbi, A. Akjouj, L. Dobrzynski, and B. Djafari-Rouhani, *Friedrich-Wintgen bound states in the continuum and induced resonances in a loop laterally coupled to a waveguide,* Physical review B **106**(12), 125414 (2022).
11. H. Friedrich and D. Wintgen, *Interfering resonances and bound states in the continuum,* Physical Review A **32**(6), 3231 (1985).
12. X.-F. Yan, X.-Y. Wang, Q. Lin, L.-L. Wang, and G.-D. Liu, *Active formation of Friedrich-Wintgen bound states in the continuum in dielectric dimerized grating borophene heterostructure,* arXiv preprint arXiv:2401.10630, (2024).
13. T. Yezekyan, S. Boroviks, O.J. Martin, and S.I. Bozhevolnyi, *Engineering quasi-bound states in the continuum in asymmetric waveguide gratings,* New Journal of Physics **26**(9), 093027 (2024).
14. S.I. Azzam, V.M. Shalaev, A. Boltasseva, and A.V. Kildishev, *Formation of bound states in the continuum in hybrid plasmonic-photonic systems,* Physical review letters **121**(25), 253901 (2018).
15. S. Joseph, S. Pandey, S. Sarkar, and J. Joseph, *Bound states in the continuum in resonant nanostructures: an overview of engineered materials for tailored applications,* Nanophotonics **10**(17), 4175 (2021).
16. S. Romano, G. Zito, S.N. Lara Yépez, S. Cabrini, E. Penzo, G. Coppola, I. Rendina, and V. Mocellaark, *Tuning the exponential sensitivity of a bound-state-in-continuum optical sensor,* Optics express **27**(13), 18776 (2019).

17. J. Wang, P. Li, X. Zhao, Z. Qian, X. Wang, F. Wang, X. Zhou, D. Han, C. Peng, and L. Shi, *Optical bound states in the continuum in periodic structures: mechanisms, effects, and applications,* Photonics Insights **3**(1), R01 (2024).
18. R. Zengerle and O. Leminger, *Phase-shifted Bragg-grating filters with improved transmission characteristics,* Journal of Lightwave Technology **13**(12), 2354 (1995).
19. H. Zhou, B. Zhen, C.W. Hsu, O.D. Miller, S.G. Johnson, J.D. Joannopoulos, and M. Soljačić, *Perfect single-sided radiation and absorption without mirrors,* Optica **3**(10), 1079 (2016).
20. R. Boidin, T. Halenkovič, V. Nazabal, L. Beneš, and P. Němec, *Pulsed laser deposited alumina thin films,* Ceramics International **42**(1), 1177 (2016).
21. Q. Zheng, X. Wang, and D. Thompson, *Temperature-dependent optical properties of monocrystalline CaF 2, BaF 2, and MgF 2,* Optical Materials Express **13**(8), 2380 (2023).
22. T. Yezekyan, V.A. Zenin, M. Thomaschewski, R. Malureanu, and S.I. Bozhevolnyi, *Germanium metasurface assisted broadband detectors,* Nanophotonics **12**(12), 2171 (2023).
23. G. Quaranta, G. Basset, O.J. Martin, and B. Gallinet, *Recent advances in resonant waveguide gratings,* Laser & Photonics Reviews **12**(9), 1800017 (2018).